\documentclass[12pt,twoside]{article}   
\usepackage[super,sort,comma]{natbib}

\usepackage{fancyhdr}		

\newcommand{\captionv}[3]{\begin{center}\parbox{#1cm}{\caption[#2]{{\sf #3}}}
        \end{center}}

\usepackage[section]{placeins}   %

\usepackage{graphicx}
\usepackage{subfigure}

\usepackage{multirow}

\makeatletter \renewcommand\@biblabel[1]{$^{#1}$} \makeatother
\newcommand{\cen}[1]{\begin{center} #1 \end{center}}

\usepackage{hyperref}
\hypersetup{ colorlinks,
    citecolor=blue,
    filecolor=blue,
    linkcolor=blue,
    urlcolor=blue
}

\usepackage{xcolor}

\definecolor{gray}{rgb}{0.6,0.6,0.6}
\definecolor{red}{rgb}{0.85,0,0}
\definecolor{green}{rgb}{0,0.85,0}
\definecolor{blue}{rgb}{0,0,0.85}
\definecolor{beige}{rgb}{0.92,0.87,0.78}
\usepackage[all]{hypcap}    

\begin{document}

\cen{\sf {\Large {\bfseries A model for studying the detectability of X-ray images using the representation of images in the primary visual cortex} \\  
\vspace*{10mm}
Antonio Gonz\'alez L\'opez} \\
Hospital Cl\'inico Universitario Virgen de la Arrixaca - IMIB. Carretera Madrid-Cartagena, 30120, El Palmar (Murcia). Spain
\vspace{5mm}\\
Version typeset \today\\
}

\pagenumbering{roman}
\setcounter{page}{1}
\pagestyle{plain}
Author to whom correspondence should be addressed. email: antonio.gonzalez7@carm.es\\
\begin{abstract}
\noindent {\bf Background:} It is well known that the visual information represented in the simple cells of the primary visual cortex V1 is spatially localized, orientation-sensitive and bandpass-filtered. In fact, it is frequently modeled as the response of Gabor filters applied to the input images. In addition, the visual information represented is subsampled, as every simple cell is not connected to all the visual receptors in the retina. On the other hand, the methods used to assess image quality of x-ray systems performing detectability tasks are usually carried out in the spatial or image domain, where it has been demonstrated that the human observer is suboptimal and is better represented by the non-prewhitening (NPW) observer.\\ 
{\bf Purpose:} This study examines the behavior of two observers, the ideal Bayesian (IBO) and the NPW, employed in the evaluation of imaging systems, in a domain that simulates the representation of visual information in V1 to verify whether the differences found in the spatial domain persist in V1 and to gain a better understanding of the degree of optimization in human observer detectability tasks.\\
{\bf Methods:} The comparison of the two observers is conducted on images of a contrast-detail phantom, which contains objects varying in size and contrast, and is performed in both the spatial and the wavelet domains, utilizing frequency bands characterized by their bandwidth and central frequency. Furthermore, the study examines the autocovariance matrices of the noise, which is the principal element that explains the differences between both observers.\\
{\bf Results:} The strong covariance exhibited in the noise among neighboring pixels in the spatial domain results in autocovariance matrices with relatively high values outside the main diagonal, leading to significant differences between the NPW and IBO observers. In the transformed domain, the band-pass decimated filters have substantially reduced the covariance between neighboring pixels, resulting in the convergence of the detectability indices of both observers.\\
{\bf Conclusions:} The vast amount of information captured by the retina, both spatially and temporally, necessitates an optimization of informational nature. This informationally optimal strategy eliminates low-frequency correlations in the processed image, after the bandpass filtering carried out by the simple cells in V1. This loss of information could contribute to the suboptimal performance of the human observer in image detection tasks. Interestingly, in  V1 the differences between IBO and NPW observers are very small. Therefore, either NPW does not accurately represent the human observer in V1, or the human observer is close to optimal in this domain.\\

\end{abstract}

\newpage     

%

\setlength{\baselineskip}{0.7cm}      

\pagenumbering{arabic}
\setcounter{page}{1}
\pagestyle{fancy}
\section{Introduction}
The human visual system processes visual information at various stages, from basic features to complex object recognition and interpretation. After receptor activation, sensory information travels coded as action potentials towards the thalamus. Then, the information travels to the cerebral cortex, in a linear and organized way as a feedforward process with consecutive hierarchical stations \cite{Cudeiro2006}. The light that reaches the retina generates electrical signals that are transmitted through the optic nerves, the optic chiasm and the lateral geniculate nucleus to the primary visual cortex (V1) \cite{Vanni2020}. V1 is responsible for edge detection, orientation and spatial frequency analysis\cite{Hubel1998}. From V1, visual information is further transmitted to visual areas V2, V3, V4 and V5, where visual perception, object recognition, motion processing, and high-level visual processing are performed\cite{Cudeiro2006,Ng2007,Vanni2020}.

Most of the optic nerve fibers end in the lateral geniculate nucleus (LGN) which forwards the pulses to V1. Therefore, most of the information transmitted to higher stages of the visual system has been previously processed in V1 \cite{Ng2007}, which makes reasonable to expect that the information necessary for object recognition, which takes place in these subsequent stages, is mainly contained within the representation of V1. For this reason, this work explores the assessment of object detectability over the type of information represented in V1, and raises the question of whether this representation should be used to evaluate imaging systems when the end-user of these systems is the human observer. It has been demonstrated that this representation results from an image processing mechanism that optimizes informational resources \cite{Barlow2012,Olshausen1997,Simoncelli2001,Lu2023}. In this processing, for example, reducing redundancy in the spatial representation of images leads to a small number of neurons encoding entire images of edge-like patterns.

From a broader perspective, V1 receptive fields, defined as regions of the image that elicit responses in one or a few neurons \cite{hubel2004,Alonso2009,Lindeberg2021}, allow the identification of the type of transformation undergone by any given image. Studies have shown that these receptive fields give rise to sparse representations, where an optimized number of resources are used to represent the huge amount of information needed to process images and movement \cite{Barlow1969}. Receptive fields of simple cells in mammalians V1 highly resemble two dimensional Gabor transforms \cite{Siegle2021,Lindeberg2021,Lu2023} and some authors have found similarities between the receptive fields of these simple cells and the basis functions of wavelet transforms \cite{Olshausen1997,Simoncelli2001,Vilankar2017}. In both cases, localized, oriented, and band-pass spatial response functions are observed  \cite{Olshausen1996}.

To assess the image quality of an X-ray imaging system, mathematical observers are utilized, which operate in the spatial or image domain. The outputs of mathematical observers are quantitative metrics that provide objective measures of image quality. Common metrics used include the signal-to-noise ratio\cite{Alvarez2014} (SNR), the contrast-to-noise ratio\cite{Desai2010} (CNR), the detectability index\cite{ICRU1995} ($d$), and the receiver operating characteristic\cite{ICRU1995} (ROC) curves. 

Image quality assessment can be carried out in the spatial domain (the representation of the image in space) as well as in linearly transformed domains \cite{Gonzalez2021} as those representing the image after the processing carried out by simple cells in V1.

In a linear system characterized by the operator $H$ and with additive noise $n$, the image $g$ of the object $f$ is obtained as \citep{ICRU1995}
\begin{equation}\label{eq:imagefor}
g=Hf+n.
\end{equation}

For an ideal Bayesian observer (IBO), decision between two hypothesis: signal present ($H_2$) and signal absent ($H_1$) should be based in the likelihood ratio $L_{IBO}$\cite{ICRU1995},
\begin{equation}\label{eq:likeratio1}
L_{IBO}=\frac{p(g|H_2)}{p(g|H_1)}.
\end{equation}

If distribution for noise $n$ is Gaussian then the decision variable $L$ can be expressed as\cite{Fukunaga1990}
\begin{equation}\label{eq:likeratio2}
L_{IBO}=(Hf_{s})^t C_{n}^{-1}g,
\end{equation}
where $Hf_s=Hf_s-Hf_{as}$ is the difference between the input signals under the two hypotheses,\cite{ICRU1995} $C_{n}$ is the autocovariance matrix for noise and $t$ indicates the transpose. 

One direct way to assess the image quality of the system is by calculating the Signal-to-Noise Ratio (SNR) of the decision variable. A higher SNR indicates a greater capacity of the system to discriminate between both hypotheses (presence and absence of the object). The SNR of $L_{IBO}$ is called detectability index $d_{IBO}$ and is calculated as \cite{ICRU1995}
\begin{equation}\label{eq:snr}
d_{IBO}^2=(Hf_{s})^t C_{n}^{-1}(Hf_{s}),
\end{equation}

If a linear transform $W$ is applied to both sides of equation \ref{eq:imagefor}, transforming the image $g$ into $Wg$, the blurred object $Hf$ into $WHf$, and the noise $n$ into $Wn$, all the previous equations can be used with the transformed variables. This is because Gaussian noise remains Gaussian after linear transformation. In particular, equation \ref{eq:snr} becomes \cite{Gonzalez2021}
\begin{equation}\label{eq:snrwl}
d_{IBO}^2=(WHf_{s})^t C_{Wn}^{-1}(WHf_{s}),
\end{equation}

In contrast to the IBO observer, which takes into account noise covariance to optimize its decision-making process, the non-prewhitening (NPW) observer neglects this noise correlation information. As a result, the NPW observer may exhibit inferior performance compared to the IBO observer, particularly in scenarios where noise correlations significantly impact the detectability of objects or features in the image. The NPW observer, when combined with the visual transfer function of the eye (NPWE) or alone, has been shown to be better than the IBO in describing human observer performance for common detectability tasks in diagnostic radiology \citep{ICRU1995,1388565,Richard2008,10.1117/12.2081655,BOUWMAN20161559}. The NPW observer decision function is described as \cite{ICRU1995}
\begin{equation}\label{eq:likeratio2}
L_{NPW}=(Hf_{s})^tg,
\end{equation}
and its detectability index as
\begin{equation}\label{eq:snrnpw}
d_{NPW}^2=\frac{((Hf_{s})^t(Hf_{s}))^2}{(Hf_{s})^t C_{n}(Hf_{s})}.
\end{equation}

.
\section{Methods}
\begin{figure}[ht]
   \begin{center}
   \includegraphics[width=0.6\textwidth]{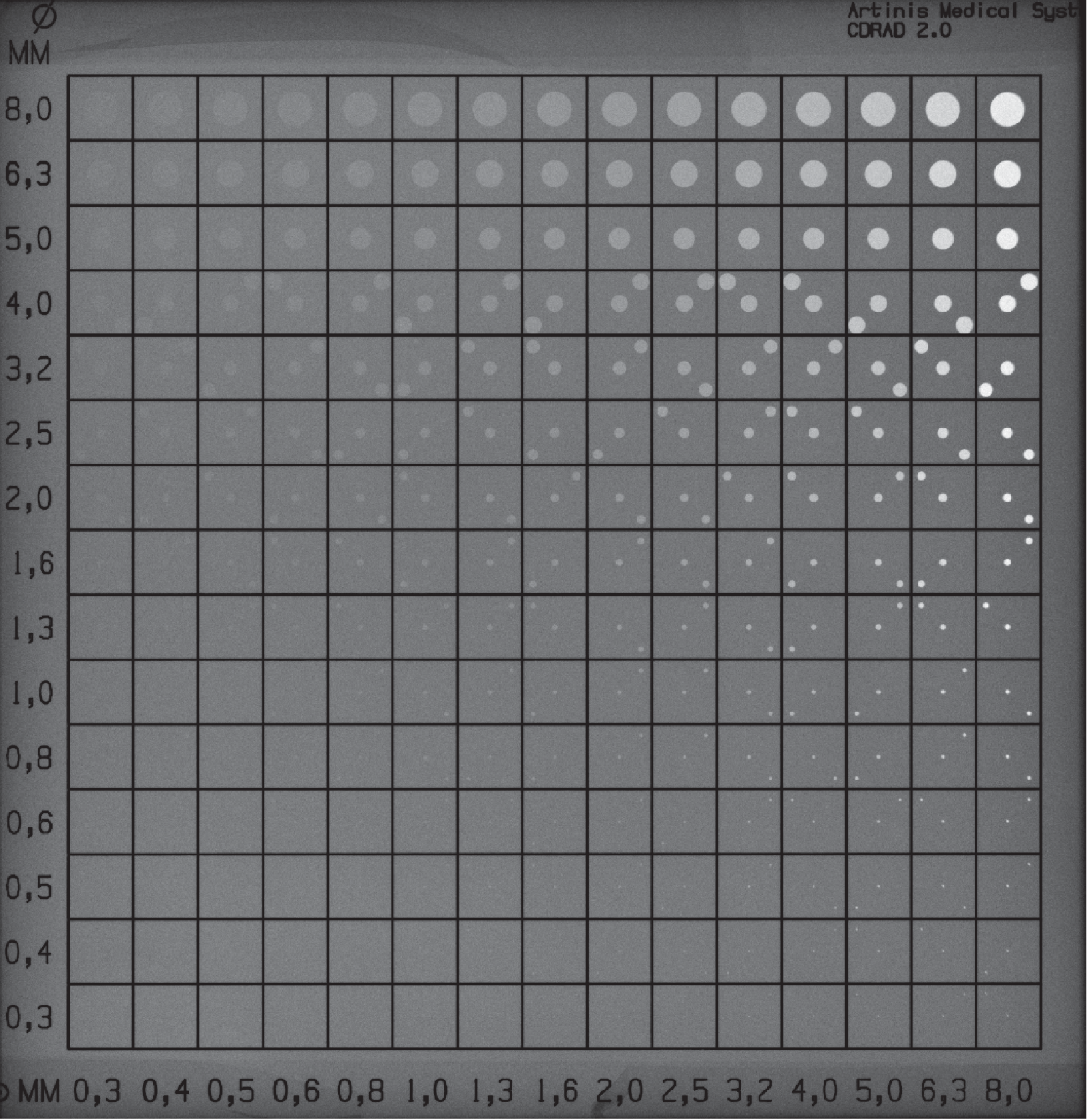}
   \captionv{12}{Short title - can be blank}
   {Average of 20 CDRAD phantom images.
   \label{fig:cdrad} 
    }  
    \end{center}
\end{figure}

The phantom reproduced in the study is the CDRAD phantom (Artinis Medical Systems), consisting of a series of cylindrical air objects in a 10 mm thick PMMA plate that provide 15 different levels of contrast and 15 different levels of resolution (figure \ref{fig:cdrad}). Different contrasts and resolutions are obtained by changing the depth and diameter of the air cylinders in the phantom form 8 to 0.3 mm.

Twenty uniform-field images were used to calculate the noise autocovariance function. The focus-to-image distance was fixed at 180 cm, employing a 70 kVp beam with filtration at the collimator exit consisting of 0.5 mm Cu + 1 mm Al, resulting in a beam quality corresponding to RQA5 \cite{IEC, aapm_tg116}. The air kerma measured at the detector entrance was 2.09 $\mu Gy$. After exposures, de-trending was carried out on the uniform images before computing noise autocovariance.
\subsection{Model template}
To perform the detectability index calculations, templates of the model were generated with the spatial resolution characteristics of the imaging system (modulation transfer function MTF) and the attenuation characteristics of the CDRAD phantom.

A 2D modulation transfer function (MTF) was constructed based on 1D MTF measurements in the two principal directions of the image \cite{IEC}. The resulting function was fitted to a double exponential, which was then employed to calculate the point spread function (PSF) of the system. Both the PSF and the cylindrical objects in the CDRAD phantom (of various diameters and heights) were sampled at a 0.01 mm interval (10 times smaller than the pixel spacing in the image detector) and convolved. After that, subsampling to 0.1 mm was carried out to generate the model templates.

To calculate the height of the cylindrical objects, the contrast produced at the imaging detector entrance $C_i$ was calculated from the differences in thicknesses of each object and the phantom's plate \cite{gonz23} for a beam quality RQA5. Then, the relationship between the air kerma beneath the object $K_s$ and beneath the plate $K_b$ was determined: $K_s = (1 + C_i)K_b$.

\subsection{Image representation in V1}
\begin{figure}[ht]
   \begin{center}
   \includegraphics[width=0.6\textwidth]{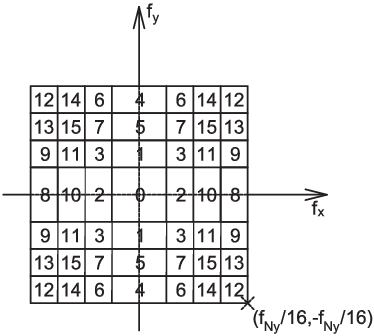}
   \captionv{12}{Short title - can be blank}
   {Frequency contents of images or nodes (2,k), k=0,1,..,15 resulting from the two-level wavelet packet carried out over the images.
   \label{fig:freccont2lev} 
    }  
    \end{center}
\end{figure}

To simulate the frequency response of the filters representing visual information in V1, horizontal, vertical, and diagonal components of a wavelet packet have been calculated for both, noise images and model templates. The wavelet packet \cite{Mallat2009} used is based on a decimated transform and employs the Haar wavelet. The number of levels in the packet is set to 2, and the analyzed nodes (see Figure \ref{fig:freccont2lev}) correspond to nodes 0, 1, 2, 3, 5, 7, 8, 10, 11, 12 and 15. These nodes contain the frequency components of the images located in 2D intervals along the horizontal, vertical, and diagonal directions of the frequency space \cite{Mallat2009}.

\section{Results}
\begin{figure}[ht]
   \begin{center}
   \includegraphics[width=0.6\textwidth]{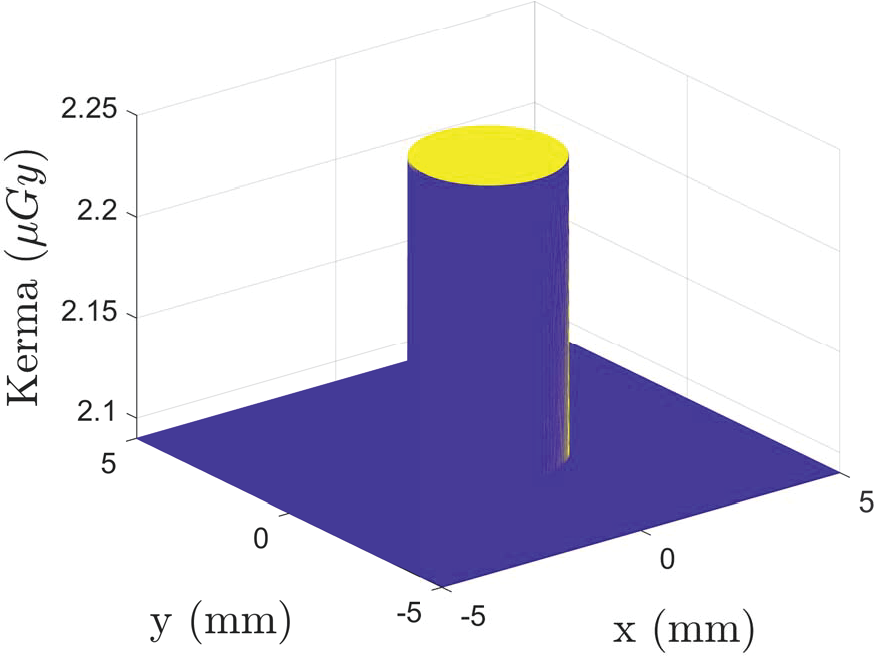}
   \captionv{12}{}
   {Cylindrical object within the CDRAD phantom, with a diameter and depth of 3.2 mm each, exposed to a radiation beam of RQA5 quality, resulting in an air kerma beneath the phantom plate of 2.09 $\mu Gy$.
   \label{fig:obj_pre} 
    }  
    \end{center}
\end{figure}

Figure \ref{fig:obj_pre} illustrates a cylindrical object with a depth of 3.2 mm and a diameter of 3.2 mm when the air kerma beneath the phantom plate is 2.09 $\mu Gy$ and the input contrast to the imaging detector is 0.0712 \cite{gonz23}. This kind of cylindrical objects are convolved with the PSF of the imaging system to create the model templates.

\begin{figure}
\centering
\subfigure[Modulation transfer function]
{\includegraphics[width=.45\textwidth]{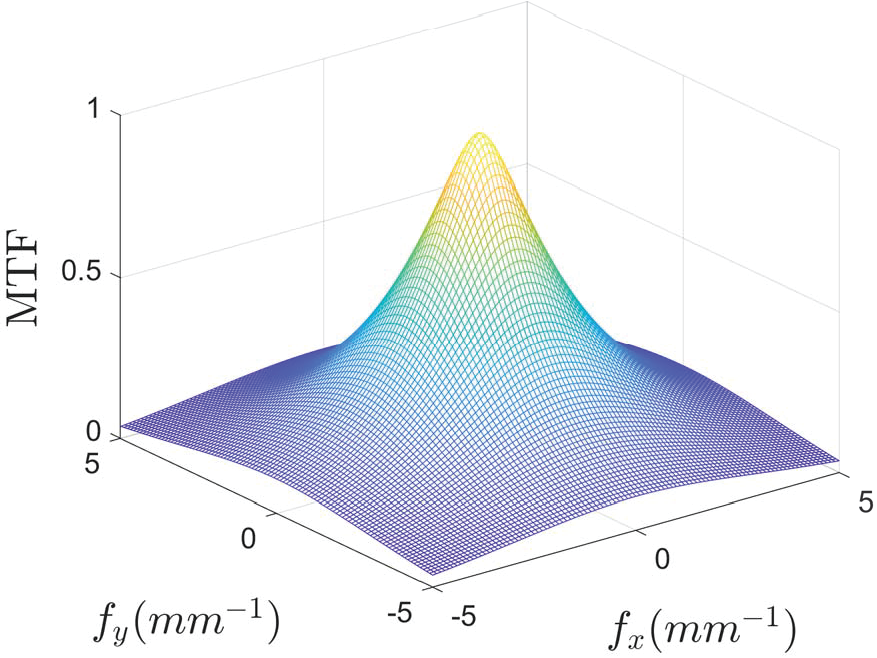}\label{fig:mtf}}
\subfigure[Point spread function]
{\includegraphics[width=.45\textwidth]{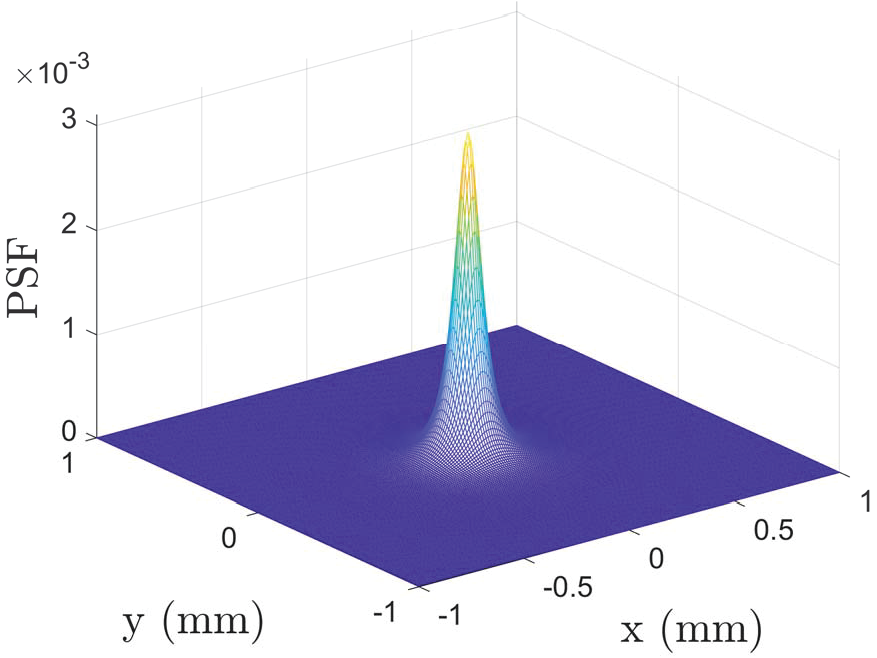}\label{fig:psf}}
\caption{Metrics for the spatial resolution of the imaging system used. From the two dimensional MTF, the PSF is calculated as its inverse Fourier transform. }
\label{fig:detail}
\end{figure}

Results of the system MTF and PSF calculations are shown in figure \ref{fig:detail}. The domain of definition for the functions has been restricted to better visualize their shapes. 

\begin{figure}
\centering
\subfigure[Templates in the spatial domain]
{\includegraphics[width=1\textwidth]{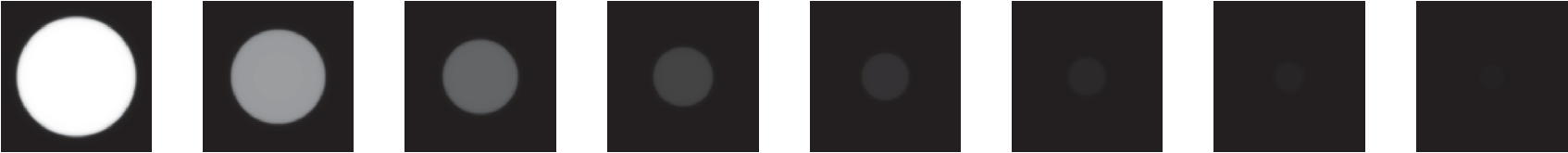}\label{fig:temp}}
\subfigure[Templates in node 10 of the wavelet domain]
{\includegraphics[width=1\textwidth]{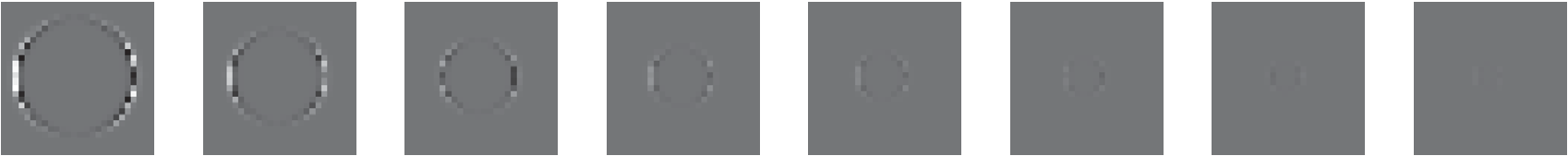}\label{fig:temp_wp10}}
\caption{Model templates used to calculate detectability indexes. The spatial templates are calculated by convolving cylindrical objects as the one shown in figure \ref{fig:obj_pre} with the PSF in figure \ref{fig:psf}, and the wavelet templates are obtained as the node 10 of the wavelet packet of the corresponding spatial template.}
\label{fig:templates}
\end{figure}

Figure \ref{fig:templates} shows the calculated templates for those objects of the CDRAD phantom with the same diameter and depth (8, 5, 3.2, 2, 1,3 0.8, 0.5 0.3 mm). Figure \ref{fig:temp} correspond to the template in the image domain and figure \ref{fig:temp_wp10} shows node 10 of the wavelet packet for the templates in the image domain.

\begin{figure}
\centering
\subfigure[Spatial domain]
{\includegraphics[width=.45\textwidth]{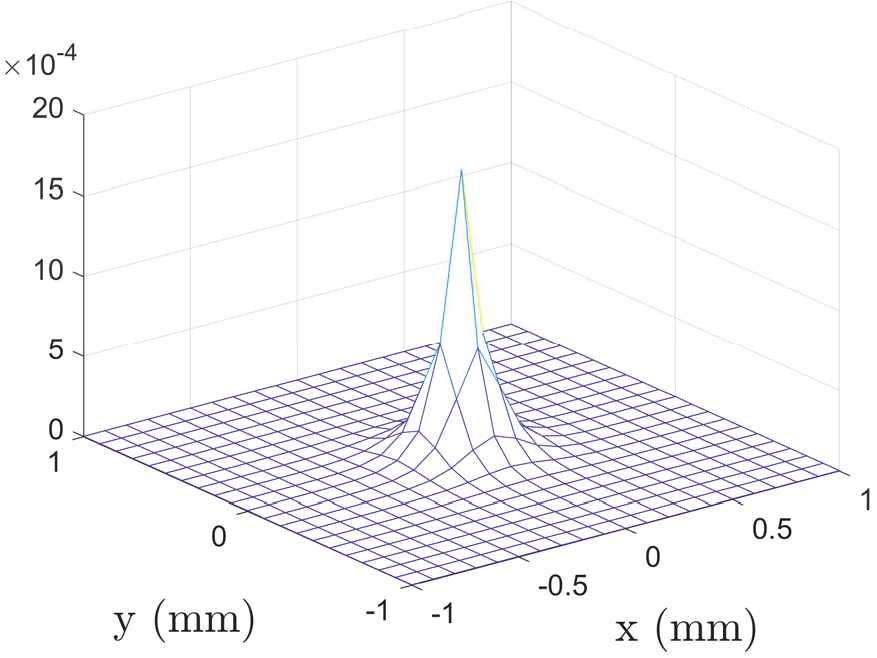}\label{fig:mac_img}}
\subfigure[Node 10 of the wavelet domain (w10)]
{\includegraphics[width=.45\textwidth]{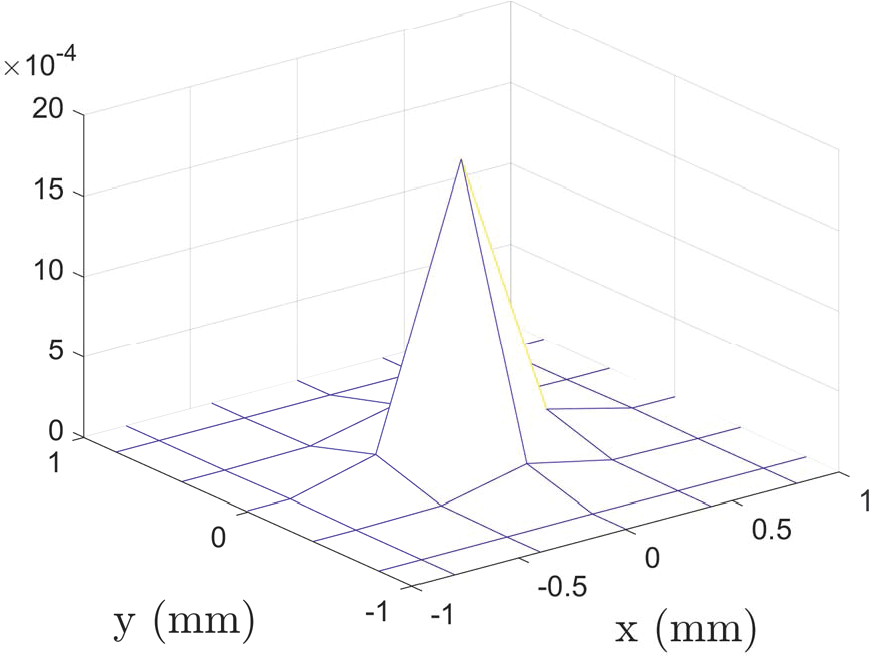}\label{fig:mac_w10}}
\caption{Autocovariance functions of noise for an entrance air kerma of 2.09 $\mu Gy$ in both, the spatial and the wavelet domains. It can be seen how, for the wavelet case, the effect of band-pass filtering and downsampling reduces the covariance values outside the origin of coordinates.}
\label{fig:macs}
\end{figure}

The autocovariance functions of noise in the spatial domain and at node 10 of the wavelet packet are depicted in Figures \ref{fig:mac_img} and \ref{fig:mac_w10}. The most significant difference between the two functions is the extent (in the discrete domain of matrix definition) of non-zero values. In the wavelet domain, the autocovariance function approximates more closely to a delta function, indicating that in this domain, the representation of noise in the image is nearly that of white noise. Conversely, in the spatial domain, adjacent pixels exhibit a pronounced statistical dependence.

\begin{figure}
\centering
\subfigure[IBO]
{\includegraphics[width=.44\textwidth]{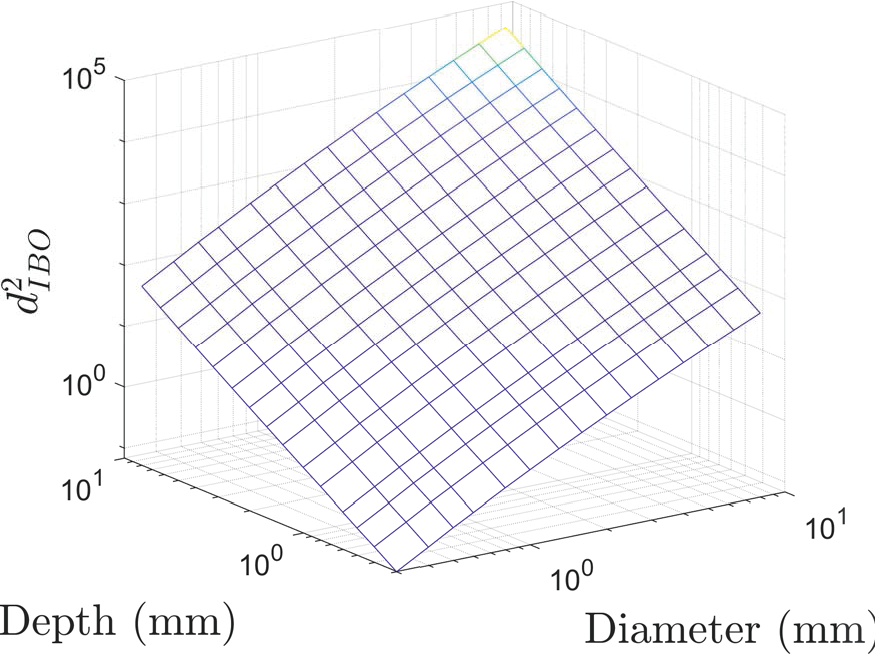}\label{fig:d2_ibo_2D}}
\subfigure[NPW]
{\includegraphics[width=.44\textwidth]{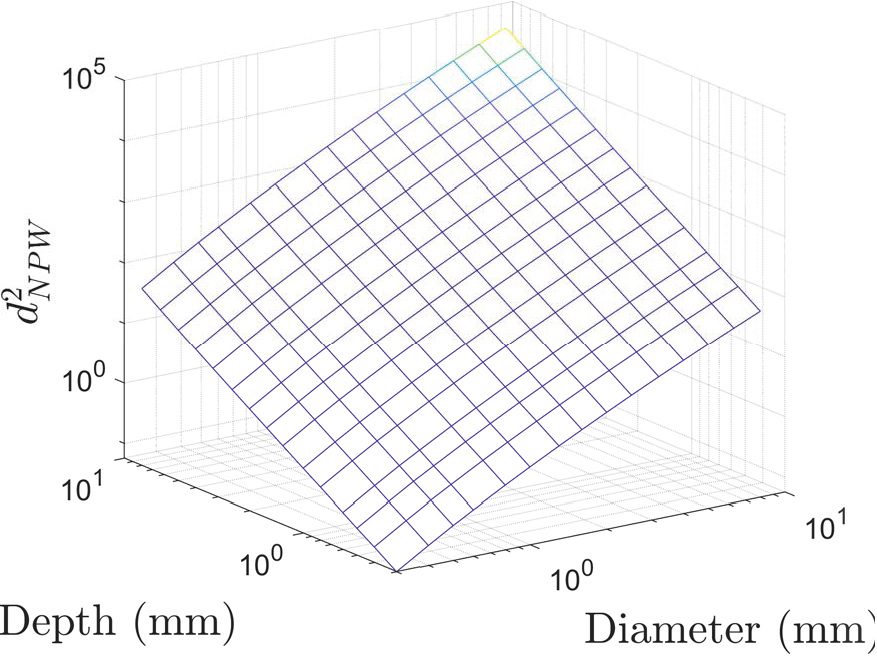}\label{fig:d2_npw_2D}}
\subfigure[$d^2$ ratio (image)]
{\includegraphics[width=.44\textwidth]{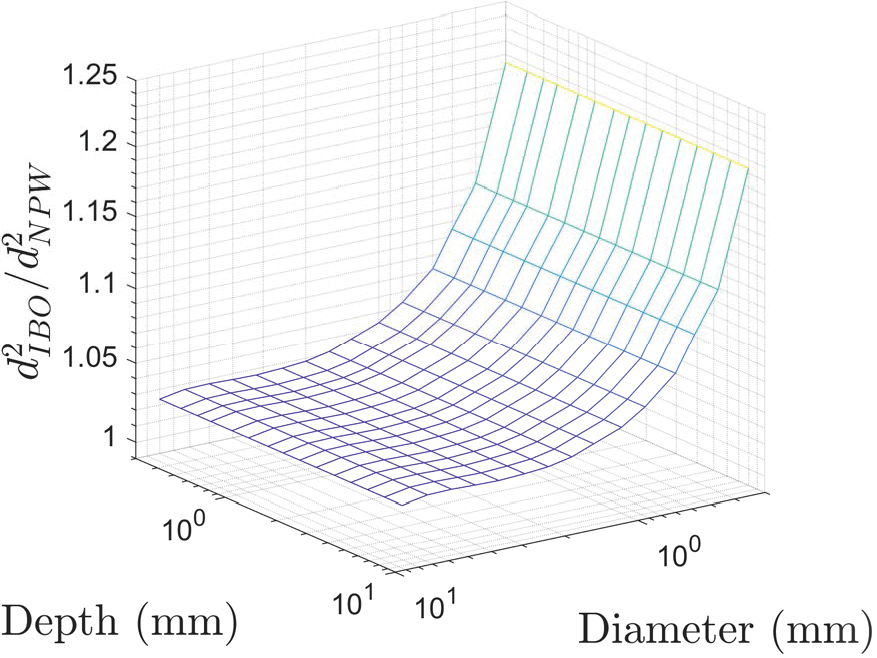}\label{fig:d2_ratio_2D}}
\caption{Detectability in the image domain for the IBO and NPW observer and ratio between IBO and NPW detectabilities.}
\label{fig:d2s}
\end{figure}

Figure \ref{fig:d2s} displays the detectability index values calculated for the two observers, IBO and NPW, in the spatial domain for the 225 objects of varying sizes and depths present in the CDRAD phantom. This figure also depicts the ratio of the results obtained by both observers (Figure \ref{fig:d2_ratio_2D}). It can be seen that, as the size of the objects decreases, the differences between both observers increase rapidly. Also, the ratio is independent of the depth of the objects. This depth is responsible for the object's contrast in the image and acts as a multiplicative factor in the Hf term of the expressions for $d^2$ in both observers (equations \ref{eq:snr} and \ref{eq:snrnpw}). As a consequence, if only the contrast is modified, $d^2$ will be multiplied by the square of this multiplicative factor in both observers, and the ratio $d^2_{IBO}/d^2_{NPW}$ will remain unaltered.

\begin{figure}
\centering
\subfigure[IBO (w10)]
{\includegraphics[width=.44\textwidth]{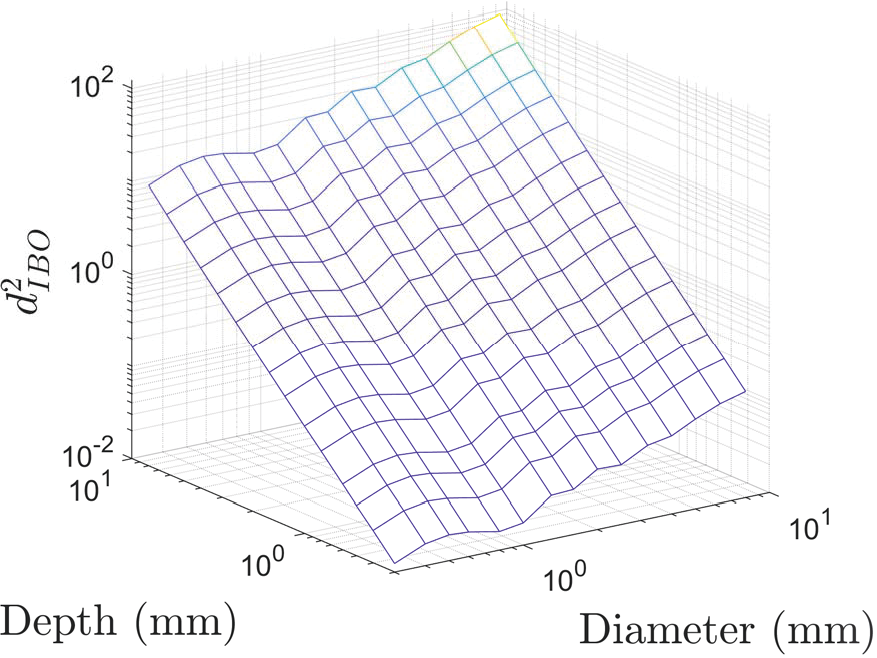}\label{fig:d2w10_ibo_2D}}
\subfigure[NPW (w10)]
{\includegraphics[width=.44\textwidth]{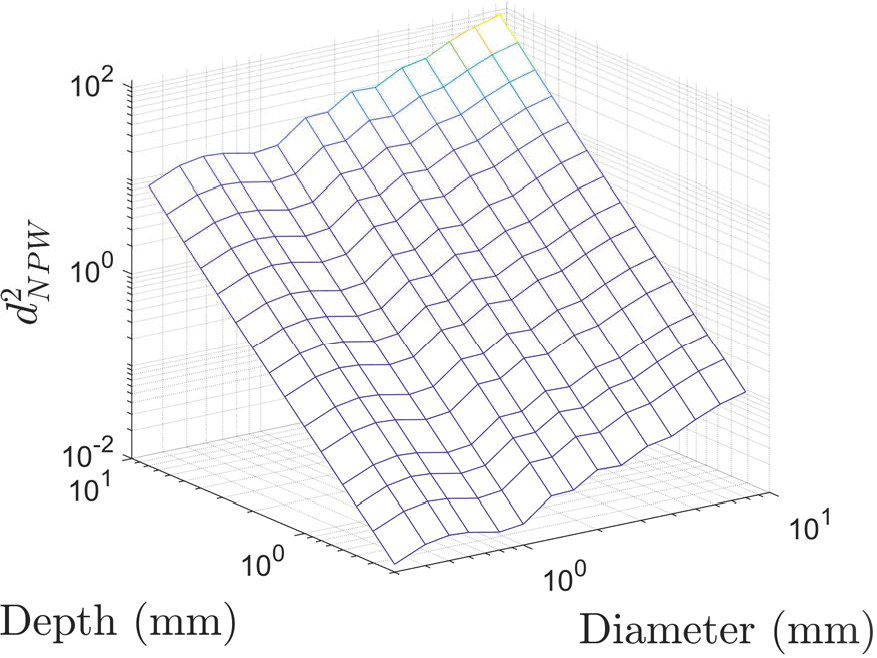}\label{fig:d2w10_npw_2D}}
\subfigure[$d^2$ ratio (w10)]
{\includegraphics[width=.44\textwidth]{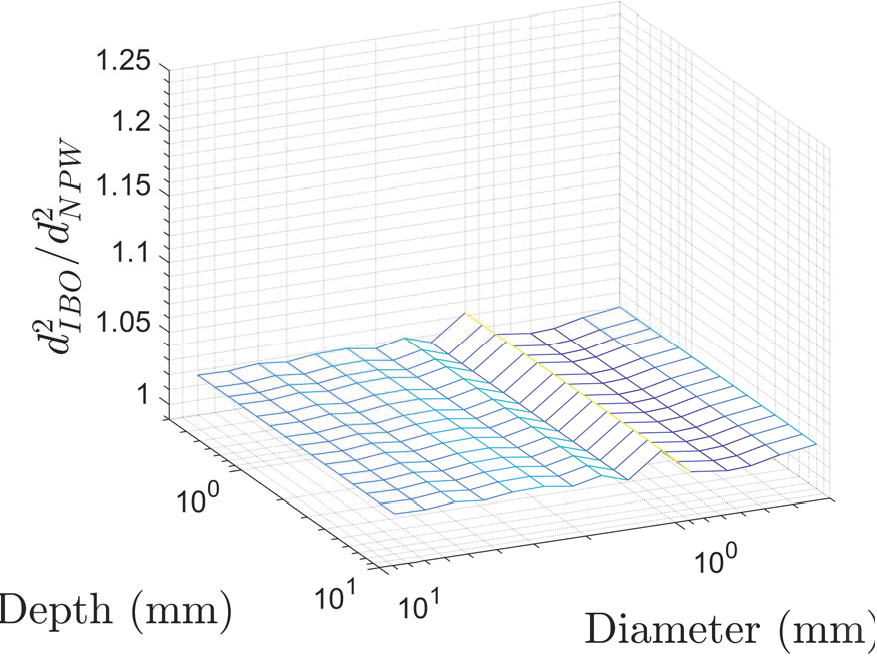}\label{fig:d2w10_ratiow_2D}}
\subfigure[$d^2$ ratio (w15)]
{\includegraphics[width=.44\textwidth]{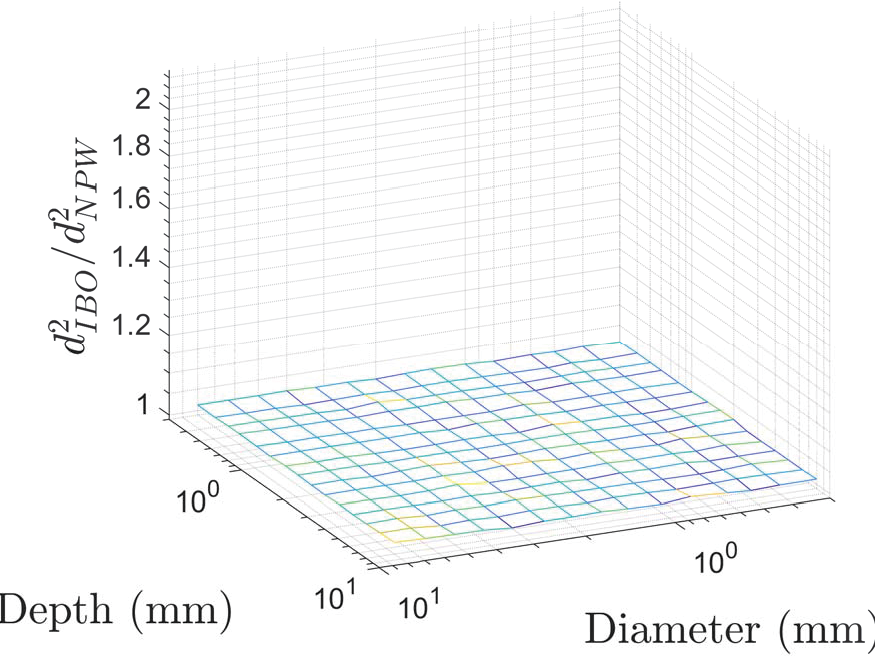}\label{fig:d2w15_ratiow_2D}}
\caption{Detectability in node 10 of the wavelet packet for the IBO and NPW observer and ratio between IBO and NPW detectabilities at nodes 10 and 15.}
\label{fig:d2sw}
\end{figure}

Detectability indexes for IBO and NPW observers at node 10 of the wavelet packet are shown in figure \ref{fig:d2sw}. Similar trends to those shown in the spatial domain can be seen in both observers. The fundamental difference in this domain is that, in this case, both observers yield very similar results regardless the size or the depth of the objects, as depicted in Figure \ref{fig:d2w10_ratiow_2D}, which illustrates the ratio between $d^2_{IBO}$ and $d^2_{NPW}$. In Figure \ref{fig:d2w15_ratiow_2D}, the same ratio is presented for node 15, revealing a similar behavior (the ratio of detectability indexes is independent of signal contrast and frequency and close to 1).

\section{Discussion}
Studies on human observer detectability have consistently noted its suboptimal performance. In tasks involving the detection of objects in image quality phantoms across various diagnostic modalities, results obtained by the ideal Bayesian observer have been significantly superior to those of the human observer.

The observed differences between the human observer and the IBO have been partially attributed to the human observer's inability to decorrelate or whiten the data, as its performance approaches that of the NPW observer, characterized by the absence of pre-whitening treatment.

The assumption of the human observer inability to pre-whiten the data appears to be in contradiction with physiological studies characterizing the human visual system. A substantial amount of information processed in the early visual system involves bandpass filtering of the image along with sub-sampling of the information available at the retina level. When both processes are combined, the result is similar to discrete wavelet analysis implemented through filter banks with bandpass filtering stages and decimation. In these cases, data decorrelation is inherent to the process, as bandpass filtering eliminates low-frequency correlations, and sub-sampling creates autocovariance matrices for the noise component that tend to be diagonal.

It is crucial to observe the role of data subsampling in reducing the width of the noise autocovariance function (Figure \ref{fig:macs}). The spatial extent of the autocovariance does not decrease in terms of physical distance in the continuous spatial domain, but it does so in the sampled or discrete domain (only the four nearest points to the origin exhibit visibly nonzero values). The justification for spatial sub-sampling in the representation of the image in V1 lies in the assumption that not all photoreceptors in the retina are connected to each of the simple cells in V1. 

The wavelet-based model employed uses critical sub-sampling, minimizing the number of samples to preserve information. In this regard, the model is optimal in eliminating redundancies in the image. This complete absence of redundancy is not necessarily a characteristic of V1 processing. In fact, it has been demonstrated that the sparse representation produced by discrete wavelet transforms gives rise to issues during information processing \cite{Selesnick2005}. For this reason, the use of overcomplete transformations is preferable: By introducing redundancy into the information representation, the quality of information processing is improved \cite{Liu2022}.

On the other hand, the poorer performance of the human observer could be partially attributed to the loss of information due to the removal of these low-frequency correlations during bandpass filtering. In most situations, the images perceived by a human observer include fairly uniform areas separated by transition zones of high intensity gradients in short distances. One question that could be formulated is if the perception of the very low frequency components of the objects, identified in the CDRAD phantom as the fairly uniform circular areas inside the disks are really used for the detection of the object when the observer is a human, or at least, to what extend does this happen?

One could argue that the suboptimal performance of the human observer in detection tasks originates from a higher-order optimization within a limited resource environment. From an evolution perspective, optimization should be one main goal of the sensory systems. However, in a limited resources scenario, informational constrains may take precedence over processing efficiency: First, we need to be able to deal with the large amount of information reaching our sensory system and then process this information in an optimal way for the required tasks.

%



%
\section{Conclusion}
To process the vast amount of visual information reaching the retina, an optimization of informational nature is required. This optimization is achieved through sparsity, wherein, after processing the image, the most relevant information becomes encapsulated within a small number of coefficients resulting from the processing. This informationally optimal strategy eliminates low-frequency correlations in the processed image and leads to a loss of optimality in detecting signals present in the images.

The findings of this work suggest that, in the transformed representation simulating information processing in V1, prewhitening is inherently carried out by bandpass filtering and decimation. In this way, detectability performances of IBO and NPW observers become very close to one another.


Furthermore, utilizing this representation points out the importance of including physiological human factors in imaging quality assessment, as the ultimate goal is to determine detectability performance when the task is carried out by human observers.

However, it is essential to acknowledge that the choice of representation depends on the specific imaging task, the intended users, and the goals of the evaluation. For certain specialized applications or automated analysis tasks, other representations or observer models might be more suitable.

\section*{Conflict of Interest}
The author has no conflicts to disclose.

\section*{Data Availability}
Data available upon reasonable request.

\section*{References}
\addcontentsline{toc}{section}{\numberline{}References}
\vspace*{-20mm}





\bibliography{./example}      



\bibliographystyle{./medphy.bst}    


\end{document}